\documentstyle[12pt,aasms4]{article}
\newcommand{\asca}{{\sl ASCA }}

\newcommand{\egret}{{\sl EGRET }}

\newcommand{\iue}{{\sl IUE }}
\newcommand{\euve}{{\sl EUVE }}

\slugcomment{}

\lefthead{Kataoka et al.}
\righthead{Variability Pattern and the Spectral Evolution of the 
BL Lacertae Object PKS 2155$-$304}
\begin{document}

\title{Variability Pattern and the Spectral Evolution of the BL Lacertae 
Object PKS 2155$-$304}
\author{J. Kataoka\altaffilmark{1,2}, T. Takahashi\altaffilmark{1,2}, 
F. Makino\altaffilmark{1}, S. Inoue\altaffilmark{3}, 
G. M. Madejski\altaffilmark{4,5},\\
M. Tashiro\altaffilmark{2}, C. M. Urry\altaffilmark{6} and 
H. Kubo\altaffilmark{7}}
\altaffiltext{1}{Institute of Space and Astronautical Science,
3-1-1 Yoshinodai, Sagamihara, Kanagawa 229-8510, Japan}
\altaffiltext{2}{Department of Physics, University of Tokyo, 7-3-1 Hongo, 
  Bunkyo-ku, Tokyo 113-0033, Japan}\
\altaffiltext{3}{Institute for Cosmic Ray Research, University of Tokyo, 
Tanashi, Tokyo 188-8502, Japan}
\altaffiltext{4}{Laboratory for High Energy Astrophysics, NASA/GSFC, 
Greenbelt, MD 20771}
\altaffiltext{5}{Dept. of Astronomy, Univ. of Maryland, College Park, MD 20742}
\altaffiltext{6}{Space Telescope Science Institute, 3700 San Martin Drive, 
Baltimore, MD 21218}
\altaffiltext{7}{Dept. of Physics, Tokyo Institute of Technology, Tokyo, 152-8551, Japan} 

\begin{abstract}
The TeV blazar PKS 2155$-$304 was monitored with the X--ray satellite 
\asca in 1994 May, as part of a multiwavelength
campaign from the radio to X--ray bands. At the beginning of the two-day
continuous observation, we detected a large flare, where the 2 $-$ 10 keV flux 
changed by a factor of 2 on a time scale of 3 $\times$ 10$^4$ sec. 
During the flare, the increase in the hard X--ray flux clearly 
preceded that observed in the soft X--rays, 
the spectral evolution 
tracking a `clockwise loop' in the flux versus photon index plane.
Ascribing the energy-dependent variability
to differential synchrotron cooling of relativistic electrons, 
we estimate the magnetic field $B$ in the emission region.
We tested two different methods of comparing the time 
series in various X--ray bands:
(i) fitting the light curves to a Gaussian and searching for the time shift 
of the peak of the flare, and (ii) calculating the
discrete correlation function.  
Both methods yielded a consistent solution of $B$ $\sim$ 0.1 G.
We also found that the flare amplitude becomes 
larger as the photon energy increases, while the duration of the flare stays 
roughly constant through the \asca energy band (0.7 $-$ 7.5 keV). 
In the framework of the time dependent synchrotron self-Compton model 
in a homogeneous region, we consider a flare where  
the maximum Lorentz factor ($\gamma_{\rm max}$) of the injected electrons 
increases uniformly throughout the emission volume. The temporal 
evolution of spectra as well as the light curves were reproduced 
with the physical parameters self-consistently determined from 
seven observables. We obtain $B$ $\sim$ 0.1 $-$ 0.2 G and a region size 
$R$ $\sim$ 10$^{-2}$ pc, for relativistic beaming with a Doppler factor of 
$\delta$ $\sim$ 20 $-$ 30. 
We discuss the significance of light travel time effects. 
\end{abstract}

\section{Introduction}
Recent observations with the \egret instrument (30 MeV -- 30 GeV; 
Thompson et al. 1993) onboard the {\sl Compton Gamma-Ray Observatory} 
({\sl CGRO}) reveal that more than 60 blazars are bright $\gamma$--ray 
emitters 
(e.g., Mukherjee et al. 1997). Multi-frequency studies of blazars show 
that the overall spectra (plotted as $\nu$$F_{\nu}$) 
have at least two pronounced continuum components: one peaking between 
IR and X--rays, and another in the $\gamma$--ray 
regime (e.g., von Montigny et al. 1995). The strong polarization observed in 
the radio and optical bands implies that the lower energy component 
is most likely produced by synchrotron radiation of relativistic 
electrons in magnetic fields, while inverse-Compton 
scattering by the same electrons is believed to be the dominant process 
responsible for the high energy $\gamma$--ray emission 
(e.g., Ulrich, Maraschi, \& Urry 1997; Ghisellini et al. 1998; 
Kubo et al. 1998). However, the source of 
the `seed' photons for the Compton process is a matter of debate. In some 
models, it is synchrotron radiation internal to the jet, as in the
Synchrotron-Self-Compton (SSC) model (Jones et al. 1974; Marscher 1980; 
K\"onigl 1981; Ghisellini \& Maraschi 1989; Marscher \& Travis 1996), 
while in others, it is radiation external to the jet, as in the External 
Radiation Compton (ERC) model (Dermer \& Schlickeiser 1993; Sikora, Begelman, 
\& Rees 1994).

Observations with ground-based Cherenkov telescopes indicate $\gamma$--ray 
emission extending up to TeV energies for 4 nearby blazars:
Mrk 421 ($z$ = 0.031; Punch et al. 1992), 
Mrk 501 ($z$ = 0.034; Quinn et al. 1996), 
1ES 2344+514 ($z$ = 0.044; Catanese et al. 1998) 
and PKS 2155$-$304 ($z$ = 0.117; Chadwick et al. 1999). 
These $extreme$ blazars are similar 
to each other in that their lower energy components peak in 
the X--ray band, while the higher energy peaks are in the TeV bands. 
Various authors have interpreted the multi-frequency spectra of TeV blazars 
by a simple homogeneous SSC model 
(e.g., Mastichiadis \& Kirk 1997; Pian et al. 1998; Kataoka et al. 1999a).

Rapid and large amplitude variability is a marked feature of blazars.
Flux variations on a time scale of hours to days indicate that blazar
emission comes from a very compact region, and is likely Doppler-boosted
towards us as a result of relativistic motion of the emitting blob of 
plasma (e.g., Mattox et al. 1993; 1997). The correlation of
variability between different bands can potentially discriminate between
various models. The multi-frequency campaign of Mrk 421, conducted in 1994
(Macomb et al. 1995; 1996; Takahashi et al. 1996) revealed correlated 
activity between the keV X--ray and TeV $\gamma$--ray emission. 
Similarly, Mrk 501 showed a remarkable flare in 1997
April, where both X--ray and TeV $\gamma$--ray flux increased by more than
one order of magnitude from the quiescent level. During the flare, the
synchrotron peak was observed around 100 keV, which is a shift by a
factor of more than 100 from its location in the quiescent level (Catanese
et al. 1997; Pian et al. 1998). For PKS 2155$-$304, TeV emission was
detected only recently (Chadwick et al. 1999).  Previous multi-frequency 
campaigns, conducted in 1991 November (Edelson et al. 1995) and 1994 
May (Urry et al. 1997), detected flares from UV to X-ray energies, 
with very different multi-frequency characteristics between the two 
epochs; there were no simultaneous TeV measurements.

From the results of those multi-frequency campaigns, many authors 
have attempted to derive the physical parameters associated with the blazar 
emission and/or the origin of the flares. For example, Takahashi et al. (1996)
associated the soft lag observed in Mrk 421 during the 1994 observation
with the energy dependence of the synchrotron cooling time, deriving
a magnetic field strength $B$ $\sim$ 0.2 G for 
$\delta$ = 5. Importantly, this was the first case where the magnetic 
field was calculated only from the observed X-ray spectral variability.
However, it is not certain that the observed time 
lag can be straightforwardly associated with the the synchrotron 
cooling time. 

Mastichiadis \& Kirk (1997) considered time-dependent 
SSC models where the variability is on time scales longer than $R/c$, 
$R$ being the size of the emission region. However, it is to be noted that the 
cooling time of the highest energy electrons, 
$t_{\rm cool}(\gamma_{\rm max})$, may well be shorter than $R/c$; in this 
case, rapid local variability could be relaxed (smoothed) 
by light travel time effects over the source on a light-crossing time scale. 
These conditions may be implied by 
the observed quasi-symmetric shape of the light curves seen both in X--rays 
and at lower energies (e.g., Urry et al. 1997; Giommi et al. 1998). 
Chiaberge \& Ghisellini (1999) took this relaxation effect into account and 
applied their model to the Mrk 421 flare in 1994. 

In this paper, we independently develop a time-dependent SSC model, to 
follow the spectral evolution and variability patterns of blazars.
In contrast to most previous works, our calculations are
$quantitatively$ compared with the observational data. We apply this model 
to the X--ray flare of PKS 2155$-$304 during the 1994 May campaign. 
Seven physical parameters for the one-zone homogeneous SSC model 
are self-consistently determined from seven observables and cross-checked 
with the magnetic field, derived directly from the `soft lag' observed in 
the X--ray band.
 
We present the \asca observation and analysis of PKS 2155$-$304 in 1994 May 
in $\S$ 2. Analyzing the time-series by various methods,
the magnetic field is deduced under 
the assumption that the energy-dependent variability is caused by
synchrotron cooling.
In $\S$ 3, we develop and summarize our time dependent model, and apply it 
to the flare in 1994 May. In $\S$ 4, we discuss the origin of 
the X--ray flare and its relation to that observed at lower frequencies. 
Finally in $\S$ 5, we present our conclusions. 

\section{Observations and Data Analysis}
\subsection{Spectral Fitting}
We observed PKS 2155$-$304 with \asca during 1994 May 19.2 $-$ May 21.3 UT, 
yielding a net exposure time of $\sim$ 80 ksec (Makino et al. 1996).
The observation was performed in a normal 1 CCD mode 
for the Solid-state Imaging Spectrometer (SIS: Burke et al. 1991; Yamashita 
et al. 1997), and a normal PH mode for the Gas Imaging Spectrometer 
(GIS: Ohashi et al. 1996). We applied standard screening procedures to the 
data and extracted source counts from a circular region centered on the target 
with a radius of 3 and 6 arcmin for SIS and GIS, respectively. 
Since the count rate of the background ($\sim$ 0.01 cts/s) and its 
fluctuation are negligible compared with the count rate from PKS 2155$-$304 in 
the 0.7 $-$ 7.5 keV band, we performed no background subtraction for the data, 
to avoid any instrumental artifacts. 

The light curve obtained from the SIS detectors is shown in Figure 1.  
The data reveal a large flare at the beginning, followed 
by lower amplitude fluctuations. This flare includes the rising and 
decaying phase;
both have a time scale of $\sim$ 3 $\times$ 10$^4$ sec. The source 
variability is somewhat different in different energy bands, and this 
is illustrated by separately plotting the light curves in the lower 
(0.5 $-$ 1.5 keV) and higher (1.5 $-$ 7.5 keV) energy X--ray bands.
Notably, the amplitude of flux change is larger at higher photon energies 
-- a factor of 2 at 1.5 $-$ 7.5 keV, while it is a factor of 1.5 in the 
0.5 $-$ 1.5 keV band -- but the flare duration seems to be nearly the same. 
Also note that the peak of the light curve in the hard X--ray band 
leads that in the soft X--ray band by $\sim$ 4 ksec (Makino et al. 1998).

The bottom panel of Figure 1 shows the time history of the hardness ratio, 
defined as the ratio of the SIS count rates at 1.5 $-$ 7.5 keV to those 
at 0.7 $-$ 1.5 keV. The hardness ratio decreases as the flux decreases, 
indicating that the X--ray spectrum tends to steepen as the source becomes  
fainter. 
This is a general feature of HBL (high-frequency peaked BL Lac objects) 
spectral variability (e.g., Giommi et al. 1990).

To investigate the rapid variability of PKS 2155$-$304, we divided the total 
exposure into 5 ks intervals.  For each segment, we analyzed the
combined SIS/GIS spectra. A single power law function and the 
photoelectric absorbing column $N_{\rm H}$ fixed at the Galactic value 
(1.77 $\times$ 10$^{20}$ cm$^{-2}$; Stark et al. 1992), often 
used to represent X--ray spectra of BL Lac objects, do not fit any of 
the spectra well. A power law fitting the data at $\gtrsim$ 2 keV was too 
steep for the data at $\lesssim$ 2 keV. We found that a single power 
law plus a free neutral absorbing column is an acceptable model, 
with reduced $\chi^2$ ranging from 0.93/dof to 1.17/dof for $\sim$ 1000 dof. 
We are aware that this is an unphysical model, 
since the source is a strong EUV emitter and the absorption cannot 
be as large as the \asca data imply. 
Nonetheless this model is a convenient representation of the curved \asca 
spectrum. 

We plot the best-fit spectral parameters for the 36 data segments in Figure 2. 
Figure 2 (a) shows the time evolution of the differential photon index in 
0.7 $-$ 7.5 keV band, while Figure 2 (b) shows the change in the 
2 $-$ 10 keV flux. The spectral shape varied such that the differential 
photon index increased from $\Gamma$ = 2.43 $\pm$ 0.02 to 3.07 $\pm$ 0.03. 
The flux also changed dramatically, with 
(14.5 $\pm$ 0.16) $\times$$10^{-11}$ erg cm$^{-2}$ s$^{-1}$ 
at the peak of the flare and 
(3.65 $\pm$ 0.06) $\times$$10^{-11}$ erg cm$^{-2}$ s$^{-1}$ 
near the end. Although we have allowed the `equivalent' photoelectric 
absorbing column $N_{\rm H}$ to be free in the model fits, it stayed nearly 
constant at $\sim$ 1.0 $\times$ 10$^{21}$ cm$^{-2}$ throughout the observation.

This spectral change is best illustrated as a correlation between the flux 
and the photon index in Figure 3, showing that the spectrum is generally
harder when brighter and steeper when fainter. Moreover the variations 
show a hysteresis, a `clockwise loop,' clearly seen 
for the flaring period, followed by gradual
steepening in the subsequent decay phase. This clockwise motion in the flux 
versus photon index plane has been seen in this particular BL Lac object 
during earlier X--ray observations (Sembay et al. 1993), and
similar behavior was clearly seen during observations of Mrk 421 in 
1994 (Takahashi et al. 1996) and H 0323+022 in 1987 
(Kohmura et al. 1994). 

\subsection{Correlation of the Variability in Various X--ray Bands}

Below, we quantify the energy dependence of the PKS 2155$-$304 
X--ray light curves, concentrating on the intense flare from 
1994 May 19.2 - 20.0 UT (indicated by arrows in Figure 1). 
 
\subsubsection{Gaussian Fit of the Light Curves}
As Figure 1 shows, the X--ray light curve during the flare of PKS 2155$-$304 
is very symmetric. This implies that a 
simple technique can be applied to measure the energy dependent 
behavior of the light curves. We fitted the light curves in various 
X--ray energy bands with a Gaussian plus constant offset function. 
In this model, the photon count rate at an arbitrary time $t$ is expressed 
as: $f(t)$ = $C_{0}$ + 
$C_{1}$ $\times$ exp ($-$($t$ $-$ $t_{\rm p}$)$^{2}$/2$\sigma^2$), 
where $C_{0}$ is a constant offset, $C_{1}$ is the 
amplitude of the flare component, $t_{\rm p}$ is the peaking time,
and $\sigma$ is the duration of the flare, respectively.

The light curves, binned at 1024 sec, were divided into 10 logarithmic-equal 
energy bands from 0.5 keV to 7.5 keV (for the SIS data) and 0.7 keV to 
7.5 keV (for the GIS data; 9 bands). 
A Gaussian provides a sufficient fit of 
the resultant energy-binned light curves. 
In Figure 4, we show examples of the Gaussian fit, which turned out 
to be a good representation of the data; 
the $\chi^2$ probability of the 
fit was $P(\chi^2$) $\ge$ 0.1 for 12 of 19 light curves. 
Note the difference in the peaking time and the difference of 
the ratio of the normalization ($C_{1}$) to the constant offset ($C_{0}$). 

Figure 5 (a) shows the dependence on energy of the flare duration derived 
from fitting the energy-binned light curves.
This result is consistent with a constant fit of 
$\sigma$ = 1.5 $\times$ 10$^4$ sec (solid line). 
Figure 5 (b) shows the energy dependence of the flare amplitude, 
defined as the ratio of the
normalization of the flare to the constant offset: 
$Ap$ $\equiv$ $C_{1}$/$C_{0}$. The flare amplitude becomes larger as the photon
energy increases. This is mostly due to the decrease in the constant 
offset ($C_{0}$) at higher energies. In other words, the spectrum during the 
flare is harder than that in the quiescent state (see Figure 2 (a)). 

In the subsequent analysis, in order to reduce errors 
and present the time lag more clearly, we binned the light curve into 
5 energy bands. 
Figure 5 (c) is the lag of the peak time, calculated from the difference 
of $t_{\rm p}$ of the Gaussian, as compared to that measured in the 3.0 $-$ 
7.5 keV band. As shown in the figure, the hard X--ray (3.0 $-$ 7.5 keV) 
variability leads the soft X--rays (0.5 $-$ 1.0 keV) by $\sim$ 4 ksec. 
We assume that the delay of the response of the soft X--ray flux 
reflects the synchrotron lifetime (cooling) of the relativistic
electrons: $t_{\rm sync}$, during which an electron loses half of its energy,
would roughly be (in the observer's frame) $t_{\rm sync}$($E_{\rm keV}$) =
1.2 $\times$ 10$^{3}$ $B^{-3/2}$ $\delta^{-1/2}$ $E_{\rm keV}^{-1/2}$,
where $E_{\rm keV}$ is the observed energy in keV 
(Takahashi et al. 1996). 
The lag is defined as the difference of $t_{\rm sync}$($E_{\rm keV}$) 
from $t_{\rm sync}$($E_{0}$), which is approximately 
$t_{\rm sync}$($E_{\rm keV}$) when $E_0$ is at much higher energies. 
Here we take $E_{0}$ to be the logarithmic mean energy in the 3.0 $-$ 
7.5 keV band. Our subsequent analysis implies $\delta$ of 20 $-$ 30, so
we use the value of 25 as a plausible value (see $\S$ 3.2).
We find the magnetic field $B$ is 0.11 $\pm$ 0.01 G for $\delta$ = 25. 
The solid line shows the model with the best fit parameter given above.

\subsubsection{Discrete Correlation Function}
We computed cross correlations using the discrete correlation function 
(hereafter DCF) of Edelson \& Krolik (1988). We divided the 0.5 $-$ 7.5 keV 
range into 5 energy bands and measured the time lag for each light curve 
compared to the 3.0 $-$ 7.5 keV light curve. The error on the lag was 
determined from the uncertainty (1 $\sigma$ error) of the peak parameter   
obtained by the minimum $\chi^2$ fitting of the distribution of the DCF 
to a Gaussian. 
The results for both SIS and GIS data are shown in Figure 5 (d). 
Again the formula for the synchrotron cooling time was used to estimate the
magnetic field strength (see $\S$ 2.2.1). The best fit value of $B$
is 0.13 $\pm$ 0.01 G for $\delta$ = 25 (solid line).

\section{Time Dependent SSC Model}
\subsection{Summary of the Numerical Code}
As we have seen in previous chapters, PKS 2155$-$304 showed rapid, 
large-amplitude variability throughout the \asca energy bands. 
To describe the data in detail,
it is essential to model the full time evolution of the 
synchrotron/Compton spectrum, incorporating radiative cooling processes, 
particle injection and escape. In the following, we consider a 
time-dependent one-zone SSC model, with a homogeneous emission region.
Steady-state one-zone models 
have been shown to provide 
a good representation of the data for similar blazars 
Mrk 421 and Mrk 501 (Mastichiadis \& Kirk 1997; Pian et al. 1998; 
Kataoka et al. 1999a).

Our kinetic code was developed as an application of the SSC model given  
in Inoue \& Takahara (1996) and Kataoka et al. (1999a), 
which is accurate in both 
the Thomson and Klein-Nishina regimes. A spherical geometry with 
radius $R$ is adopted. Our code is qualitatively 
similar to the one given in Chiaberge \& Ghisellini (1999); in 
particular, we consider both the time evolution of the electron and photon 
distributions, as well as the effects introduced by the different 
photon-travel times in different parts of the emission region. For the steady 
state solution of the SSC code, we compared our results with Band \& Grindlay 
(1985; 1986). The time evolution of the 
synchrotron spectrum was also compared with the analytic solutions given by 
Kardashev (1962), for the simplest case of synchrotron cooling. 
The kinetic code will be described in detail in a forthcoming paper 
(Kataoka et al. 1999b); here we briefly summarize the calculations,
and present the results as applied to PKS 2155$-$304. 

We first consider a given electron population characterized by 
$N_e(\gamma,t)$, where $\gamma$ is the electron Lorentz 
factor and $N_e$ is the electron number density per unit volume per unit 
energy. The time $t$ can be regarded as an initial time for the evolution of 
the electron and photon spectra.

Given the electron population, the synchrotron emission including 
self-absorption can be calculated using the spherical solution to 
the radiative transfer equation,
\begin{equation}
L_{\rm sync} (\nu, t) = 4 \pi^2 R^2 \frac{j_\nu}{\alpha_\nu} 
(1 - \frac{2}{{\tau_{\nu}}^2}[1 - e^{ - \tau_{\nu}}(\tau_{\nu} + 1 ) ] )  ~,
\end{equation}
where $j_{\nu}$ and $\alpha_{\nu}$ 
are respectively the emission and absorption coefficients for synchrotron 
radiation (e.g., Blumenthal \& Gould 1970; Gould 1979; Inoue \& Takahara 
1996). The optical depth in the plasma cloud along 
the central line of sight is $\tau_{\nu}$ = 2$\alpha_{\nu}$$R$. 

We calculate the inverse Compton emission incorporating the effects of 
cross section reduction in the Klein-Nishina regime. The differential photon 
production rate $q(\epsilon,t)$ is
\begin{equation}
q(\epsilon, t) = \int d \epsilon_0 n(\epsilon_0, t)\int d \gamma 
N_e(\gamma, t) C(\epsilon, \gamma, \epsilon_0) ~,
\end{equation}
where $\epsilon_0$ and $\epsilon$ are respectively the soft photon energy 
and the scattered photon energy in units of $m_e c^2$. 
The number density of soft photons per energy interval is
$n(\epsilon_0,t)$,
and $C(\epsilon,\gamma,\epsilon_0)$ is the Compton kernel of Jones (1968).
As the optical depth for Compton scattering is small, 
the self-Compton luminosity is obtained from the relation
\begin{equation}
L_{\rm SSC} (\nu, t) = \frac{16}{3} \pi^2 R^3 j_\nu^{\rm SSC} ~ ,
\end{equation}
where $j_\nu^{\rm SSC}$ is the emission coefficient of the Compton emission,
related to $q(\epsilon,t)$ as 
$j_\nu^{\rm SSC}$ = $h$$\epsilon$$q(\epsilon,t)$(4 $\pi$)$^{-1}$ and 
$h$ is the Planck constant. 

The time evolution of the high-energy electrons in the magnetic field and
the photon field is described by the following kinetic equation:
\begin{equation}
\frac{\partial N_e(\gamma,t)}{\partial t} = \frac{\partial}{\partial \gamma}
[(\gamma^{\cdot}_{\rm sync} + \gamma^{\cdot}_{\rm SSC}) N_e(\gamma,t)] + 
Q(\gamma,t) - \frac{N_e(\gamma,t)}{t_{\rm esc}} ~ ,
\end{equation}
where $Q$ and $t_{\rm esc}$ are the injection rate and the escape time of 
the electrons, respectively. For simplicity, we set $t_{\rm esc}$ to a 
constant value. 
The synchrotron and SSC cooling rates for a single electron are 
expressed as 
\begin{equation}
\gamma^{\cdot}_{\rm sync} = \frac{4 \sigma_T \gamma^2 U_B}{3 m_e c} ~,
\hspace{10mm}\gamma^{\cdot}_{\rm SSC} = \int \epsilon d \epsilon \int d 
\epsilon_0n(\epsilon_0, t) C(\epsilon, \gamma, \epsilon_0) ~,
\end{equation}
where $U_B$ is the magnetic field energy density. Thus the right-hand side 
of equation (4) is now described by the quantities at time $t$, and solved 
numerically to obtain the electron population at time $t$+$\Delta$t. 
We adopted an implicit difference scheme by Chang \& Cooper (1970) to obtain 
non-negative and particle number conserving solutions. The iteration of the 
above prescription gives the electron and photon populations at an arbitrary 
time. 

Finally, we consider light-travel time effects. We proceed under the 
assumption that electrons are injected uniformly throughout a homogeneous 
emission region. Such a description may be adequate as long as the injection 
timescale is  longer than $R/c$ (Dermer 1998), but is unphysical for shorter 
timescales since the particle injection process itself should take (at least)
 $\sim$ $R/c$ to influence the whole region. In a more realistic picture, a 
thin shock front may propagate through the emission region with a finite 
velocity $v_{\rm s}$, supplying freshly accelerated electrons only 
in the front's vicinity (Kirk, Rieger \& Mastichiadis 1998). 
However, such a detailed 
calculation necessarily involves some additional, uncertain parameters, not 
to mention the assumption of a particular geometry. Instead, we will choose 
the injection to be uniform over the emission volume.
Such a choice, albeit unrealistic, will allow us to clarify the 
important role of light 
travel time effects on blazar variability (We will, however, choose the 
duration of the injection to be comparable to $R/c$). The observer would see, 
at any given time, photons produced in different parts of the source, 
characterized by an electron distribution of a different age. The observed 
spectra must then be a sum of the corresponding different spectra. Below, we 
will argue that light travel time effects may be essential in interpreting 
the shapes of blazar light curves.

In order to incorporate the light-travel time effects properly in the 
calculation, we divide the source into (2$t_{\rm crs}$/$\Delta$$t$) 
slices of $\Delta$$R$ thickness for each, where $t_{\rm crs}$ (= $R/c$) is 
the source light-crossing time and $\Delta$$t$ is the time-step of 
the calculation. 
The interval $\Delta$$t$ must be shorter than the shortest relevant time 
scale, e.g, the synchrotron or Compton cooling times or the 
injection time scale. We define $\Delta$$R$ $\equiv$ $c$$\Delta$$t$. 
The schematic view of this division of the emission blob into slices 
is given in Figure 6.  
We  first consider the slices in the source frame $K'$, with line 
of sight perpendicular to the surface of the slices. 
In the observer's frame $K$, emission will be concentrated in the forward 
direction  within a narrow cone of half-angle 
1/$\Gamma$  (e.g., Rybicki and Lightman 1979), if the blob moves with a 
Lorentz factor $\Gamma$ ($\simeq$ $\delta$ for our case). 

In our current assumption, the synchrotron cooling occurs homogeneously
over the emission region, because 
the electrons are injected uniformly into the entire volume
with a constant magnetic field strength.
Even so, the Compton scattering should not be uniform, 
as each electron should experience the changes in the photon field 
after a different time interval depending on its position. 
The synchrotron photons should require a time $\sim$ $R/c$ 
to be completely scattered, causing an additional delay for the 
response of the Compton photon spectrum.
However, for this particular TeV 
blazar, synchrotron cooling must dominate over Compton cooling 
for the energy loss process of the electrons, as 
(1) the synchrotron luminosity is greater than the Compton luminosity, and 
(2) the reduction of the cross section in the Klein-Nishina regime
significantly decreases the Compton scattering efficiency (see also $\S$ 3.3). 
In addition, the present gamma-ray data is not sampled well enough
to allow a comparison to the Compton model spectrum on very short timescales.
Therefore,
in calculating the Compton spectrum and energy loss,
we approximate that the synchrotron radiation 
(i.e., the soft photons for inverse Compton scattering) instantaneously 
fills the whole emission region, reducing the computational 
time (see also Chiaberge \& Ghisellini 1999).

\subsection{Self-Consistent Solution for PKS 2155$-$304} 
To reproduce the flare of PKS 2155$-$304 in 1994 May, we first model the 
multi-frequency spectrum from the radio to TeV bands, and determine the physical 
parameters relevant for the quiescent emission. Since the source activity is 
well-sampled, and appears to be rather smooth, we consider the level 
at $\sim$ MJD 49492 as a $locally$ quiescent state (see Figure 1). 
We use an accurate solution of the one-zone homogeneous SSC model ($\S$ 3.1), 
to describe the multi-frequency spectrum of PKS 2155$-$304. 
To specify the spectral energy distribution for this source, 
we need the magnetic field $B$, region size $R$, beaming factor $\delta$, 
escape time $t_{\rm esc}$, and the electron injection spectrum 
as input parameters. 
We adopted a specific form for the latter, $Q(\gamma)$ = 
$q_e$ $\gamma^{-s}$ exp($-$$\gamma$/$\gamma_{\rm max}$), 
where $\gamma_{\rm max}$ is the maximum Lorentz factor of 
the electrons. For the minimum Lorentz factor of the electrons, 
$\gamma_{\rm min}$, we set $\gamma_{\rm min}$ = 1, 
although this choice is not important for our results.
Thus seven free-parameters are required to specify the model.

All these parameters are associated with seven observables:
synchrotron peak frequency, $\nu_{\rm s}$;
Compton maximum frequency, $\nu_{\rm c}$;
synchrotron break frequency, $\nu_{\rm br}$;
variability time scale, $t_{\rm var}$;
synchrotron energy flux, $f_{\rm sync}$;
Compton energy flux, $f_{\rm SSC}$;
and radio (millimeter) spectral index, $\alpha$.
Despite the equal numbers of 
observables and parameters (7 observables for 7 parameters), the model cannot 
be specified uniquely, because the region size is described by an inequality 
$R$ $\le$ $c$$t_{\rm var}$$\delta$. 
To determine model parameters $uniquely$, we make the approximation 
$R$ $\sim$ $c$$t_{\rm var}$$\delta$ in the following. 

The parameters for the one-zone homogeneous SSC model are characterized by 
these observables as follows 
(see also Mastichiadis \& Kirk 1997; Tavecchio, Maraschi \& Ghisellini 1998). 
The peak frequency of the synchrotron component in the observer's frame is 
given as
\begin{equation}
\nu_{\rm s} \simeq 1.2 \times 10^6 B \delta \gamma_{\rm max}^2 ~.
\end{equation}

We define the Compton maximum frequency $\nu_{\rm c}$ as the frequency 
where the Compton luminosity decreases by an order of magnitude  
from its peak observed flux (assumed here the highest energy bin in the 
\egret data). For our PKS 2155$-$304 data, this corresponds to  
$\sim$ 1 TeV and coincides with the VHE range (Chadwick et al. 1999).
At this highest photon energy ($\sim$ $h$$\nu_{\rm c}$), the emission from the 
TeV blazars are probably suppressed by the decreased Compton scattering 
cross section in the Klein-Nishina regime. Thus we obtain the approximate 
equality of the maximum electron energy and the photon energy 
\begin{equation}
\gamma_{\rm max} m_e c^2 \delta \simeq h \nu_{\rm c} ~.
\end{equation}

The third relation is obtained from the ratio of the synchrotron luminosity 
to the Compton luminosity (see Kataoka et al. 1999a):
\begin{equation}
u_B = \frac{{d_L}^2}{R^2 c \delta^4} \frac{{f_{\rm sync}}^2}{f_{\rm SSC}} ~,
\end{equation}
where $d_L$ is the luminosity distance. Combining equations (6) $-$ (8), 
we can express $R$ as a function of 
only one parameter, $\delta$. By equating $R$ with 
$c$$t_{\rm var}$$\delta$, we can determine the relation between 
observables and $\delta$, $R$, $B$ and $\gamma_{\rm max}$ uniquely. 

To determine the escape time $t_{\rm esc}$, another relation is required.
An injection of a power law energy distribution of electrons up to a certain 
maximum energy ($\gamma_{\rm max}$) into the radiating region should yield a 
steady-state electron distribution with a break in its index at a 
characteristic energy ($\gamma_{\rm br}$), determined by the balance between 
radiative cooling and advective escape (e.g., Inoue \& Takahara. 1996). 
Thus at $\gamma$ = $\gamma_{\rm br}$, we expect 
$t_{\rm esc}$ = $t_{\rm cool}$. This can be written as 
\begin{equation}
t_{\rm esc} = \frac{3 m_e c}{4 (U_B + U_{\rm soft}) \sigma_T \gamma_{\rm br}} ~,
\end{equation}
where $U_{\rm soft}$ is the soft photon density, which is related to $U_B$ 
via $U_B$/$U_{\rm soft}$ = $f_{\rm sync}$/$f_{\rm ssc}$ in the Thomson 
regime. Thus we can express $t_{\rm esc}$ as a function of $B$, $\delta$, and 
$\nu_{\rm br}$. $\nu_{\rm br}$ is determined as the frequency where the 
spectral index starts to deviate from that in the radio (millimeter) band,
which presumably corresponds to the uncooled portion of the electron distribution. 

We summarize the `input' observables in Table 1, and `output' physical 
quantities in Table 2. The relation between the observables and model 
parameters is also shown in Table 2.
The remaining parameter $q_e$ for 
injected electrons was determined to agree with the synchrotron luminosity 
$f_{\rm sync}$. The slope of the injected electrons $s$ was determined 
simply by 2$\alpha$ + 1. As a result, we obtain a magnetic field $B$ = 0.14 
G, region size $R$ = 7.7 $\times$ 10$^{-3}$ pc, and beaming factor 
$\delta$ = 28. This result is quite similar to that reported in 
Tavecchio, Maraschi \& Ghisellini (1998). 
They investigated the allowed parameter space for ($B$, $\delta$), and 
conclude that $B$ $\simeq$ 0.2 G with $\delta$ $\simeq$ 25 is 
the most probable parameter set for PKS 2155$-$304.

To estimate the uncertainties of the parameters listed in Table 2 and the 
resultant model prediction of the multi-frequency spectrum, 
we considered different cases where the selected value of one observable 
is varied while the other six observables remain unchanged (see Table 1). 
We first changed the relation $R$ = $c$$t_{\rm var}$$\delta$ to 
$R$ = (1/3) $c$$t_{\rm var}$$\delta$. In this case, we obtain 
$B$ = 0.19 G and $\delta$ = 37 with region size $R$ = 1.0 $\times$ 
10$^{16}$ cm. Next we increased the Compton energy flux $f_{\rm ssc}$ by 
a factor of 3. This results in $B$ = 0.12 G, $\delta$ = 25 and $R$ = 
2.1 $\times$ 10$^{16}$ cm.  Finally  we increased the maximum Compton 
frequency by a factor of 2. This changes the result most significantly, 
where $B$ = 0.05 G, $\delta$ = 40 and $R$ = 3.3 $\times$ 10$^{16}$ cm.  
However, when $\delta$ $\gtrsim$ 30, the TeV flux becomes considerably 
higher than that observed, because most of the Compton scattering takes 
place in the Thomson regime.  
Also note that any change of $\nu_{\rm br}$ keeps all the parameters unchanged 
except for $t_{\rm esc}$.

To obtain the steady state solution for electron and photon spectra based on 
the parameters listed in Table 2, we start from the initial condition 
$N_e(\gamma, 0)$ = $Q(\gamma)$ and calculate the 
time evolution of the spectra to more than $t$ $>$ 30 $t_{\rm crs}$, 
assuming constant injection and escape. In Figure 7, 
we show the multi-frequency spectra of PKS 2155$-$304, 
as well as the calculated SSC model lines. The solid lines in Figure 7 show 
the synchrotron self-Compton spectra at the `steady state', based on the
parameters in Table 2. Although most of the data in Figure 7 were obtained 
non-simultaneously, variability is not large except for the X--ray band, 
so these spectra provide a reasonable constraint on the source parameters. 
One can see the discrepancy in the radio band, but at the higher  
frequencies, the model represents the observed spectrum quite well. 
The discrepancy in the radio band can be an effect of 
the emission arising from larger distances than the location of the keV/TeV 
emitting region. The one-zone models cannot account for this low-frequency 
emission, which is thought to be produced in a much larger region of the 
source (e.g., Marscher 1980).

\subsection{Modeling of the PKS 2155$-$304 Flare in 1994} 
Based on the physical parameters selected above, and the resultant `steady 
state' spectrum as an initial condition, we model the X--ray flare of 
PKS 2155$-$304 in 1994 May. Various types of flaring behavior 
were investigated by 
changing the parameters for the injected electron spectrum and/or the physical 
quantities in the emission region. In the following, we simulated the light 
curves, as well as the time evolution of spectra, when $\gamma_{\rm max}$ 
increased by a factor of 1.6 during one $t_{\rm crs}$ interval from the 
start of the flare. After the calculation was performed in the source 
frame, it was transformed into the observer's frame for comparison with the 
observational data. We take the time-step to be $\Delta$$t$ = 2000 sec 
throughout the calculation. This corresponds to $\Delta$$t$ $<$ 100 sec 
in the observer's frame, much shorter than the variation time scales, 
such as the synchrotron cooling time (see $\S$ 2).
 
We varied the injected electron spectrum as $Q_e(\gamma)$ = 
$q_e$ $\gamma^{-s}$ exp($-$$\gamma$/1.6$\gamma_{\rm max}$) for 
0 $\le$ $t$ $\le$ $t_{\rm crs}$ and $Q_e(\gamma)$ = $q_e$ 
$\gamma^{-s}$ exp($-$$\gamma$/$\gamma_{\rm max}$) otherwise.
In Figure 8, we show the calculated light curves from \euve to \asca energies. 
The flux was normalized to that for the steady state ($t$ = 0), and the time 
axis was normalized to the source light-crossing timescale ($R$/c). 
The symmetric light  curve during the flare is reproduced quite well. 
Notably, the peaking time of the flare at lower energies lags behind that 
for higher energies, and the 
amplitude of the flare becomes larger as the photon energy increases,
which agree qualitatively with the observational data.

The position of the peak time is determined by the balance of slices 
in which the emitted flux is increasing and slices in which the decaying 
phase has already started. For example, at the highest X--ray energy band 
which corresponds to $\gamma$ $\sim$ $\gamma_{\rm max}$, the peak of the 
light curve will occur at a time $t$, where  $t_{\rm crs}$ $<$ $t$ $<$ 
2$t_{\rm crs}$. 
This is because these electrons in the slices cool much faster than $t_{\rm crs}$ 
and the volume which contains the flare information is maximum at the center 
of the sphere, which can be observed after $t_{\rm crs}$  from the start of 
the flare (see Figure 6).
The whole emission region requires  2 $t_{\rm crs}$ to be completely 
visible to the observer. 
This will cause a decaying of the flux 
(foreside slices) and an increasing of the flux (backside slices) at the same 
time. However, in the lower energy band where 
$t_{\rm cool}$ $\gg$ $t_{\rm crs}$, the electrons do $not$ cool effectively. 
The emitted flux will continue to increase even after 2 $t_{\rm crs}$. 
This combination of the increasing/decreasing 
phase of slices can cause a time lag in the position of the peak as can be 
seen in Figure 8.
 
The dashed lines in Figure 2 (a) and (b) show the spectral evolution 
calculated from the adopted model. The 2 $-$ 10 keV flux was obtained by 
integration of the calculated spectrum in the 2 $-$ 10 keV band, so it can 
be readily compared with the observational data. The photon spectral index 
was simply determined from the ratio of the fluxes at 0.7 keV to 7.5 keV. 
Since the \asca observation started after the onset of the flare, we shifted 
our simulated light curves in the time-axis for comparison. 
The fact that the peaking time of the 
photon index leads that for the flux is quite well reproduced. 
This can be also seen in Figure 3, where the `clockwise' hysteresis 
in the flux (2 $-$ 10 keV) versus photon index (0.7 $-$ 7.5 keV) plane 
can be clearly seen.

To make a quantitative comparison of the observed and modeled 
light curves in different energy bands, we analyzed the light curves in the 
same way as that for the observational data. The dashed lines in Figure 
5 (a) $-$ (c) were calculated from a Gaussian fit of the simulated light 
curves, converted to the observer's frame. Figure 5 (a) shows the 
duration of the flare, determined from the standard deviation ($\sigma$) 
of the Gaussian. This stays nearly constant at $\sim$ 1.5 $\times$ 10$^{4}$ 
sec, but the model shows a sign of increase at lower X--ray energies 
-- $\sigma$ for 0.5 keV is 8 $\%$ longer than that for 5.0 keV. 
Although we do not detect any such increase 
(broadening) in the observed X-ray light curves (Figure 1), 
flares observed by \euve and 
\iue showed considerably longer time scales than that in the \asca 
band (Urry et al. 1997). This may suggest that the duration of the flare 
actually increases at lower energies, if both flares have the 
same origin.  Also note that the Gaussian fit of the simulated light curves 
could make systematic errors on $\sigma$, if the symmetry of the light 
curves is broken only in the lowest X-ray energy bands 
(see also the discussion in $\S$ 4).

The amplitude of the flare ($Ap$ $\equiv$ $C_1$/$C_0$) was 
calculated to be 0.5 for 0.5 keV and 1.6 for 5 keV -- precisely in agreement 
with the observational data (Figure 5 (b)). We compared the light curves 
at 5 keV to those for the lower energies, and found a `soft-lag' 
where the 0.5 $-$ 1 keV photons lag behind 5 keV photons about 4 ksec, 
quantitatively in agreement with the observational data (Figure 5 (c)). 
Importantly, this result implies that the observed time lag is well 
represented by the difference of the synchrotron cooling time scales. 

We also computed the discrete correlation function for the model light 
curves. The result is given in Figure 5 (d). Again, we obtained a similar 
result as with Gaussian fits, and verified a `soft-lag' where the 
0.5 $-$ 1 keV photons lag behind 5 keV photons about $\sim$ 4 ksec. 

Finally, in Figure 7 we show the time evolution 
of the multiwavelength synchrotron self-Compton spectra 
after the start of the flare, at $t$/$t_{\rm crs}$ = 1.0, 1.6 and 2.4 
(for comparison, see also Figure 8). 
The X--ray spectrum clearly becomes harder when 
the flux increases. Also note that the flux variation of the Compton 
spectrum is smaller than that for synchrotron spectrum. This is because the 
reduction of Compton cross section in the Klein-Nishina regime strongly 
suppresses the increase of the flux at the highest photon energies 
($\sim$ $h$$\nu_{\rm c}$). 

We also investigated other scenarios for the flare, such as increasing the 
normalization of the injected electrons ($q_e$) or the magnetic field $B$ 
strength, but these did not give good representations of the data. 
In a more realistic situation, all of these parameters probably vary 
simultaneously, but our modeling implies that the most essential parameter 
during the flare is $\gamma_{\rm max}$. 
We will discuss more about this in the next chapter. 

\section{Discussion}
Our calculation shows that a one-zone homogeneous SSC model incorporating 
time evolution of the electron and photon distributions can be a viable tool 
to study the rapid variability of blazars. 
Investigating both the light curves and the spectral evolution of
the X--ray flare of PKS 2155$-$304 in 1994, 
our modeling reproduces the observational data quite well, 
despite the relatively simple assumptions regarding the flare mechanism. 

All the physical parameters for the model were self-consistently determined 
from seven observables (Table 1). We found that the magnetic field 
$B$ = 0.14 G, region size $R$ = 7.7 $\times$ 10$^{-3}$ pc, and beaming 
factor $\delta$ = 28 give a good representation of the observational 
data (Table 2 and Figure 7). Although the present spectrum 
is too sparsely sampled, especially in the hard X--ray and $\gamma$--ray bands, 
we showed that 
our derived parameters are not affected by more than a factor of 2 
even if taking the uncertainty of the observables into account. 

The magnetic field derived from the multiband spectrum is nicely consistent 
with that calculated from the X--ray variability based on the 
synchrotron cooling model ($\S$ 2); 
while the Gaussian ($\S$ 2.2.1) and DCF ($\S$ 2.2.2) fits 
give slightly different values for $B$, both yielded the same value within 
the uncertainties, $B$ $\sim$ 0.1 G. Regarding this point, we should note
that the errors on the lag determined from the DCF are somewhat ambiguous, 
because they are calculated from the errors on the peak of the 
distribution of DCF. When the DCF distribution is not well 
represented by a Gaussian, this can cause additional errors in 
$B$, which may account for the difference in the magnetic field 
determined from the Gaussian fit of the light curves.

We also found that in the X--ray bands, the duration of the flare is expected 
to be nearly constant in energy, forming very symmetric time-profiles. 
We considered a model where rapid local variability is relaxed (smoothed) 
on a light crossing time scale (e.g., Chiaberge \& Ghisellini 
1999; Giommi et al. 1998). In this scenario, the rising and falling phases of 
the flare are characterized by the same time scale $R/c$, forming a 
quasi-symmetric shape of the light curves. This can be clearly seen in 
Figure 1 and Figure 4, where the observed light curves are well represented 
by a simple Gaussian function (see $\S$ 2.2.1). The standard deviation 
$\sigma$ determined from the Gaussian fit (see Figure 5 (a)) 
is thus expected to be on the order of $\sim$ $R$/$c$$\delta$, in the 
observer's frame. Importantly, the synchrotron cooling time is calculated 
to be $\sim$ 7 $\times$ 10$^3$ sec for 1 keV photons (see $\S$ 2.1.1) 
which is shorter than $\sigma$ (assuming $B$ = 0.11 G and 
$\delta$ = 25). However, at lower energies where $t_{\rm sync}$ 
$\ge$ $R$/$c$$\delta$, it is possible that the symmetry of the light curves is 
broken and a longer time scale decay could be observed -- 
this is why the model line in Figure 5 (a) begins to rise in the lowest 
X--ray energy bands ($\lesssim$ 1 keV).

Effects other than that due to light travel time 
could also make the light curves symmetric, having constant duration
independent of the photon energy. For example, Dermer (1998) considers 
a similar but simpler model, assuming that the electron injection event
continues for at least a light-crossing time.
In describing the time evolution of the electron/photon spectra, 
he discusses quantities which are averaged over timescales shorter than $R/c$. 
Highlighting the case of an $s$ = 2 electron injection power-law index, 
he found that the light curves can become symmetric, 
with energy-dependent lag in the mean time of 
the flare in various energy bands. Importantly, he showed the flare mean time 
stays nearly constant in energy if $t_{\rm cool}$ is less than or comparable to 
the duration of injection, as is the case for our modeling. 
However, it is to be noted that symmetric light 
curves can be reproduced in his model 
only when the particular electron injection distribution of $s$ = 2 is selected 
(see equation (2) of Dermer 1998).

We believe that relaxing of rapid, local variability 
by light-travel time effects must generally be important in blazar variability.
As noted above, if the only timescales relevant in the light curves 
are those associated with cooling and injection,
we should mainly observe $asymmetric$ light curves, 
except for the special case of $s$ = 2. However, flares with 
symmetric time-profiles were observed many times from this particular source 
previously, in various states of activity with X-ray differential photon 
indices ranging from 2.0 to 3.5 (e.g., Sembay et al. 1993; 
Edelson el al.1995; Giommi et al. 1998; Chiappetti et al. 1999),
perhaps reflecting a range of electron injection indices.
To account for the general variability trends of this source,
it thus seems preferable to include light-travel time effects. 
Also note that symmetric light curves are not unique phenomena for 
PKS 2155$-$304, but also commonly found for other TeV blazars 
(e.g., Takahashi et al. (1999) for Mrk 421).

To describe the PKS 2155$-$304 flare in 1994 May, we adopted a model where 
the flare is due to an increase of $\gamma_{\rm max}$
which continues for one source light-crossing time scale, 
$t_{\rm inj}$ = $t_{\rm crs}$. 
Regarding this point, we note that while longer durations of electron injection, 
i.e., $t_{\rm inj}$ $\gg$ $t_{\rm crs}$, may be physically possible, 
this cannot have been the case for our PKS 2155$-$304 flare. 
When the injection is much longer than $t_{\rm crs}$, 
both the electron and photon distributions will have 
enough time to reach a 'new' equilibrium state, forming a $plateau$ in the 
light curve (e.g., Mastichiadis \& Kirk 1997; Chiaberge \& Ghisellini 1999). 
Such a $plateau$, although not as common, has occasionally been
observed in the case of Mrk 421 (see, Takahashi et al. 1999;  Kataoka et al. 
1999b for more detailed discussion), but not in our PKS 2155$-$304 data. 
If the injection results from a shock propagating with velocity $v_{\rm s}$
through the emission region, 
the duration of the injection event should be characterized by 
the shock crossing timescale $R/v_{\rm s}$ 
(e.g., Kirk, Rieger \& Mastichiadis (1998)). Thus our current assumption of 
$t_{\rm inj}$ = $t_{\rm crs}$ corresponds to a highly relativistic shock, 
$v_{\rm s} \sim c$.

Despite the simple description of the injection 
(an increase of $\gamma_{\rm max}$ with other parameters fixed), 
it provides a good representation of the observational data. 
More realistically, this may correspond to a new population of
electrons injected with a harder distribution during the flare.
In the picture of shock acceleration, 
electrons with larger $\gamma$ require a longer time to be accelerated 
(e.g., Kirk, Rieger, \& Mastichiadis. 1998),
so our assumption of an increase in 
$\gamma_{\rm max}$ can be a good approximation 
as long as the acceleration timescale $t_{\rm acc}$ at $\gamma_{\rm max}$
is well shorter than $t_{\rm cool}$.

Our numerical calculation also predicts the flare amplitude, 
duration, and the time lag of the same flaring event in the \iue and \euve 
bands. The agreement with the data (Urry et al. 1997) is qualitatively
correct but quantitatively there are discrepancies of a factor of $\sim$4
if we interpret these flares as having the 
same origin. The time lags from the highest \asca band ($\sim$ 7 keV) 
are estimated to be $\Delta$$t_{EUVE}$ = 0.2 day and $\Delta$$t_{IUE}$ = 0.5 
day, compared to the observed values of $\sim$1~day and 
$\sim2$~days, respectively.
However, the \euve `peak' is very flat, and both \euve and \iue flares
could conceivably represent multiple injections.
The flare amplitude is estimated from the model to be $Ap$ ($\equiv$ 
$C_{1}$/$C_{0}$; see $\S$ 2.2.1) = 0.2 for \euve (compared to $\sim0.5$ 
observed) and $Ap$ = 0.05  for \iue (compared to $\sim0.35$ observed).
The duration of the flare is calculated to be 0.2 
day for \euve and 0.5 day for \iue. These values are nearly consistent with 
\euve observational data, but disagree more strongly with \iue data. 
These discrepancies may indicate that another mechanism is required to 
account for the low-energy flare, or that there are multiple smaller flares 
uncorrelated with X-ray variations, or that the one-zone homogeneous 
assumption can not be applied at lower energies.

Recent work by Georganopoulos \& Marscher (1999) also find it difficult
to reproduce the flare amplitude observed with \iue in 1994 May, even
assuming a more realistic, inhomogeneous SSC jet model with 
many parameters. They suggest that one of the origins of the discrepancy 
could be a mild re-acceleration of the electrons in the injected plasma. 
Such a mild re-acceleration may be manifested mainly at lower energies, 
thereby increasing the flare amplitude relative to that at higher energies.
However, the short coverage of the \iue, \euve and \asca data during 
the 1994 May campaign is insufficient to deduce further conclusions. 

\section{Conclusions}
The TeV blazar PKS 2155$-$304 was observed with the X--ray satellite \asca in
1994 May, as part of a multiwavelength campaign. A rapid, 
large-amplitude flare was detected at the beginning of the observation. 
During the flare, the change in the hard X--ray flux
led the change in the soft X--ray flux by $\sim$ 4 ksec, as derived by 
two different methods. We showed that the light curves can be
fitted with a Gaussian plus constant offset. 
The magnetic field $B$ was estimated to be $\sim$ 0.1 G for $\delta$ = 25, 
assuming the energy-dependent delays are due to the different 
synchrotron lifetimes of the relativistic electrons.
We also found that the flare amplitude increased as the photon energy 
increased, while the duration was nearly constant throughout the \asca band. 
Using a time dependent model of synchrotron 
self-Compton emission in a homogeneous region, we considered a scenario
where the electrons are uniformly injected into the emission region, 
and rapid, local variability is smoothed out by light-travel time effects. 
In our modeling, the lag of the peaking time can be explained by the 
different combination of the increasing/decreasing phase of local
emission regions (``slices'').
The spectral energy distribution of PKS 2155$-$304 from the radio to TeV bands 
was well represented by the parameters $B$ $\sim$ 0.1 $-$ 0.2 G, $R$ $\sim$ 
$10^{-2}$ pc and $\delta$ $\sim$ 20 $-$ 30. Importantly, 
these parameters were self-consistently determined from seven observables. 
We found that the X--ray flare in 1994 May 
is well explained by a change in the maximum energy of the electron Lorentz 
factor ($\gamma_{\rm max}$) by a factor of 1.6. 
However, the relation to the \iue and \euve flares remains 
uncertain because of the short coverage of the observations. 
More data, especially longer and continuous coverage 
at neighboring frequencies such as UV to X--rays, 
are necessary to fully understand this source, 
and the physical processes operating in blazars.

\acknowledgments
We would like to thank Dr. J. G. Kirk and Dr. M. Kusunose for their 
helpful discussion on the time-dependent SSC model. We also thank an anonymous 
referee for his constructive comments and discussion.
This research has made use of the NASA/IPAC Extragalactic Database (NED) 
which is operated by the JET Propulsion Laboratory, Caltech, under 
contact with the national Aeronautics and Space Administration.
J.K acknowledges the Fellowships of the Japan Society for 
Promotion of Science for Japanese Young Scientists, 
Grant-in-Aid for Scientific Research No. 05242101,
and Grant-in-Aid for COE Research No. 07CE2002 by the 
Ministry of Education, Culture and Science, Japan.

\clearpage

\figcaption[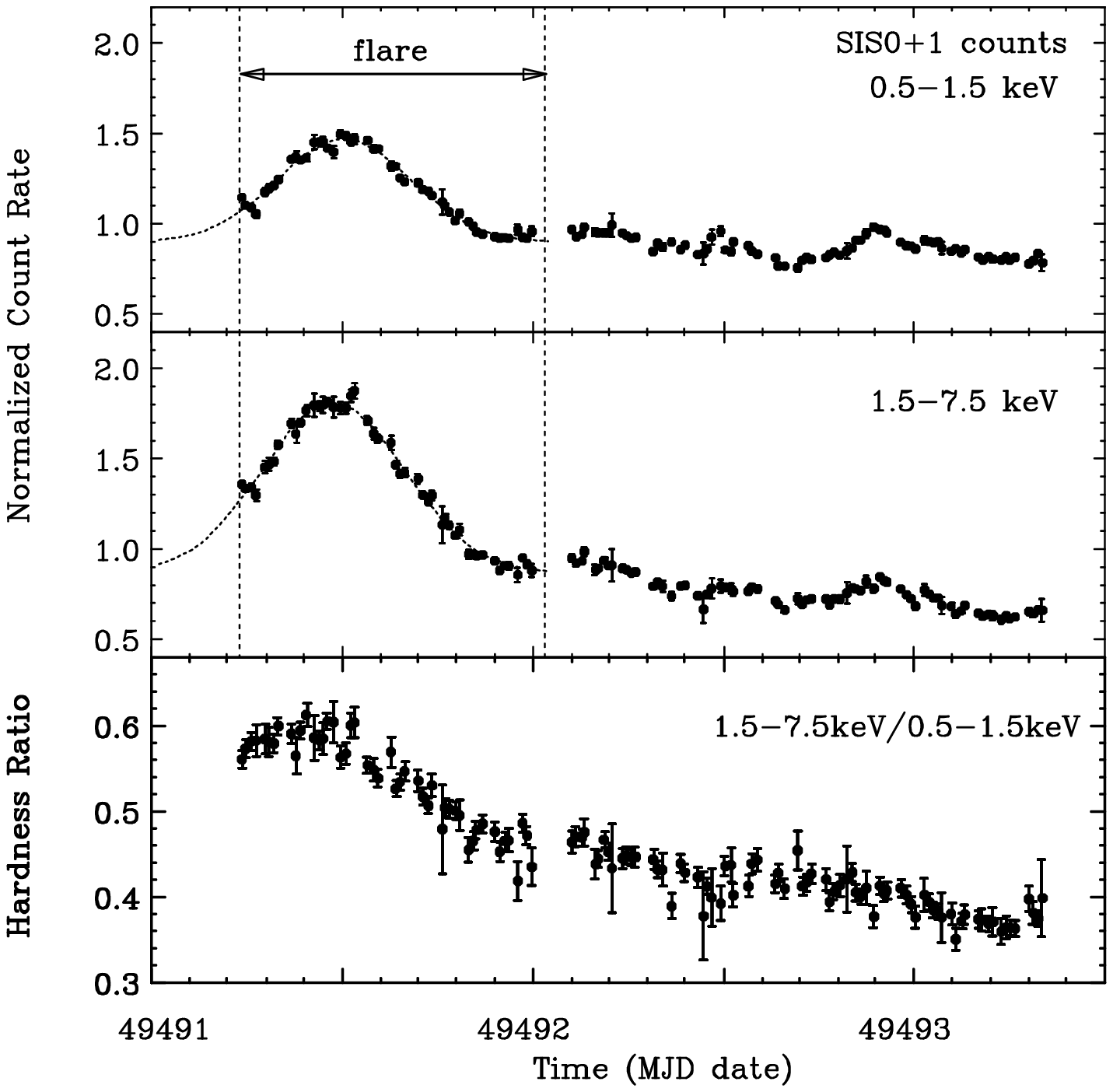]{Time history of the X--ray emission of 
PKS 2155$-$304 during the 1994 May campaign. 
Upper Panel: Combined SIS light curves in the lower X--ray 
band (0.5 $-$ 1.5 keV). 
Middle Panel : Combined SIS light curves in the the higher X--ray band 
(1.5 $-$ 7.5 keV).  Both light curves are normalized to their average count 
rate. The arrows indicate the time interval used for the Gaussian fit and 
discrete correlation function (DCF) in $\S$ 2. The best fit Gaussian is 
superposed on each curve to guide the eye.  
Lower Panel: Time history of the hardness ratio, defined as the 
ratio of the SIS count rates at 1.5 $-$ 7.5 keV to those at 0.7 $-$ 
1.5 keV. \label{fig.1}}

\figcaption[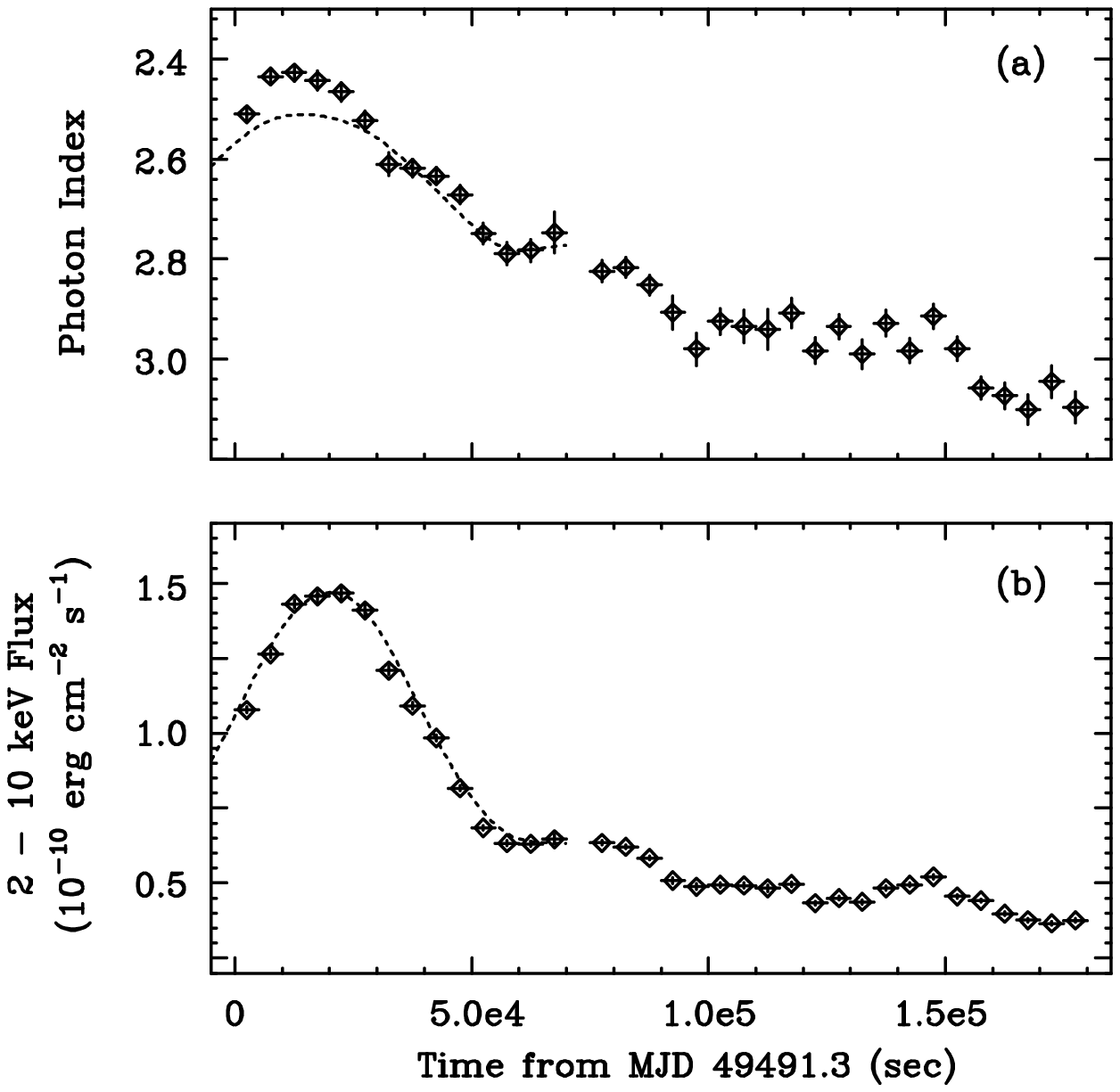]{Detailed time history of PKS 2155$-$304 during the 1994 
May campaign. Each data point corresponds to an equal 5 ksec interval and 
all SIS/GIS data are combined for the fit. The model is a power law with 
free absorption. (a): Variation of the differential photon index in the 0.7 $-$ 
7.5 keV band. The dashed line is a model prediction as described in $\S$ 3. 
(b): Variation of the 2 - 10 keV flux in units of 10$^{-10}$ erg cm$^{-2}$ 
s$^{-1}$. The dashed line is a model prediction discussed from 
$\S$ 3.\label{fig.2}} 

\figcaption[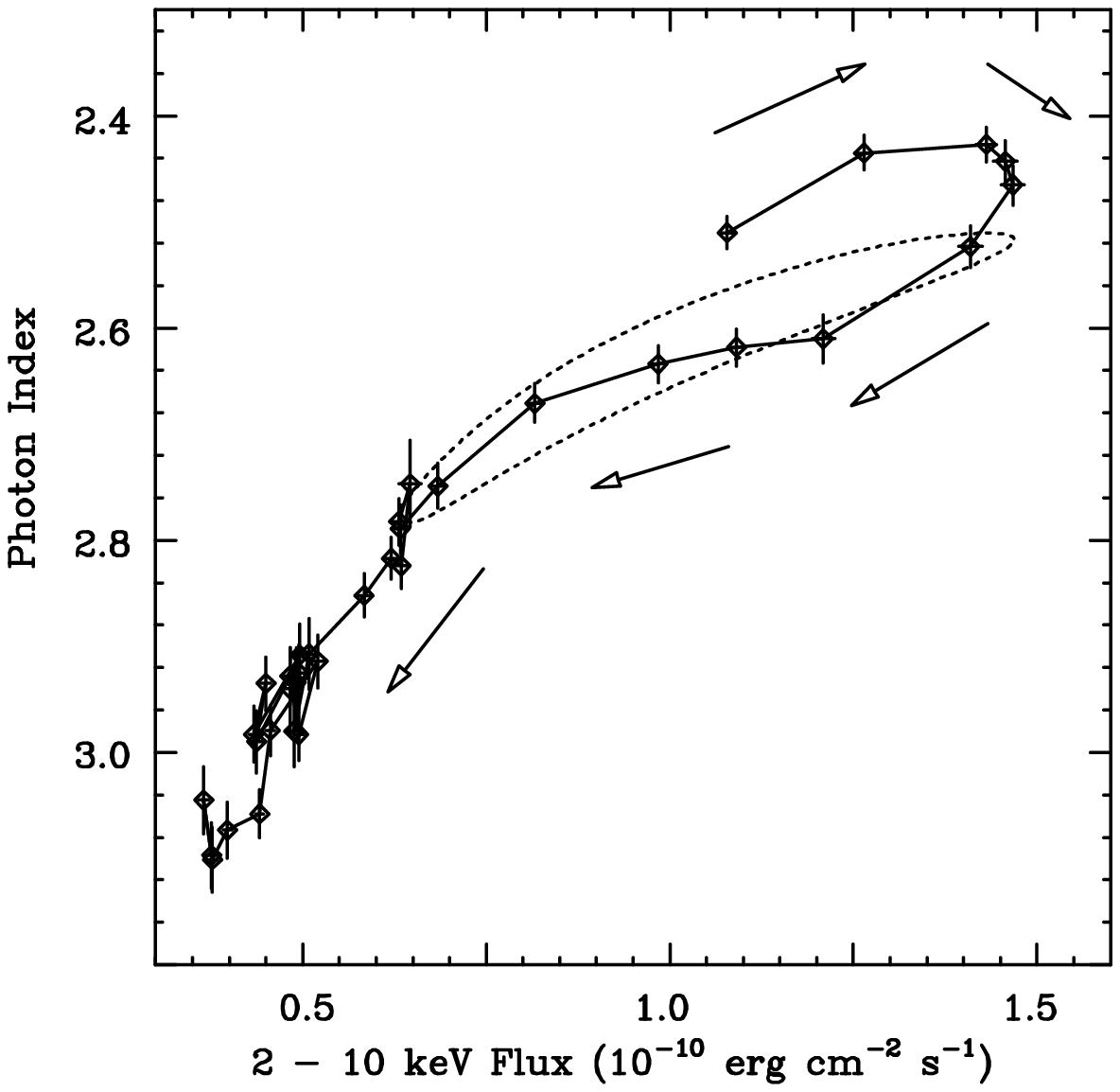]{Evolution of the X--ray spectrum of PKS 2155$-$304, 
in the flux versus photon index plane. 
Arrows indicate the evolution during 1994 May observation. 
A `clockwise loop' 
is clearly seen. The solid line connects the observational data, 
while the dashed line is the model prediction given in $\S$ 3.\label{fig.3}}

\figcaption[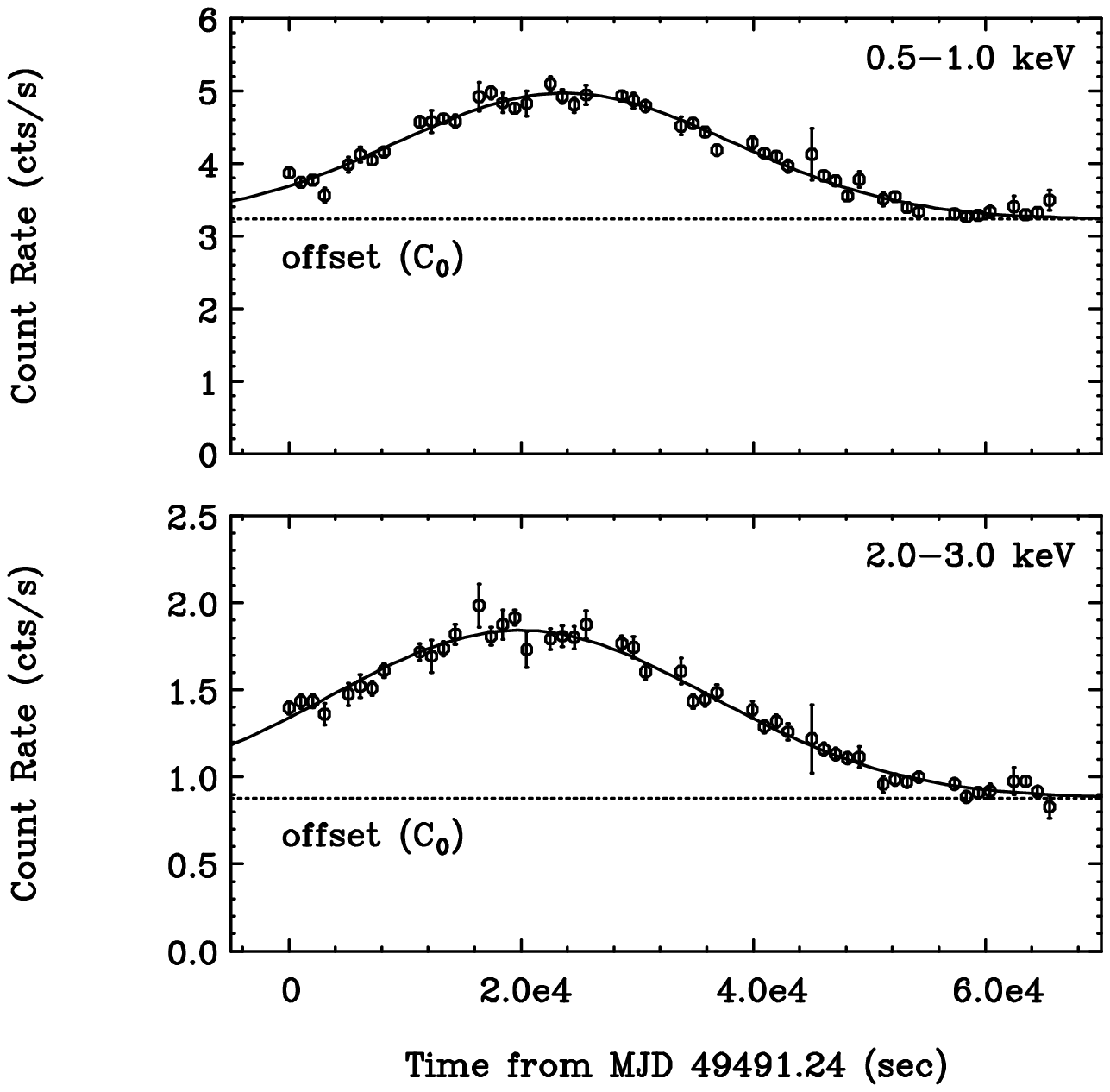]{A Gaussian fit to the time profile of PKS 2155$-$304  
during the flare (May 19.2 $-$ 20.0 UT; see also Figure 1). 
The count rates of both SIS detectors are summed. 
The solid line corresponds to the best fit 
with $f(t)$ = $C_{0}$ + $C_{1}$ $\times$ 
exp ($-$($t$ $-$ $t_{\rm p}$)$^{2}$/2$\sigma^2$), 
while the dashed line is the constant offset $C_{0}$ (see $\S$ 2.2.2). 
Upper Panel: Light curve in the 0.5 $-$ 1.0 keV band; $\chi^2_{\rm red}$ = 
1.5 for 49 dof. Lower Panel: Light curve in the 2.0 $-$ 3.0 keV band; $\chi^2_{\rm red}$ = 1.1 for 49 dof.\label{fig.4}}

\figcaption[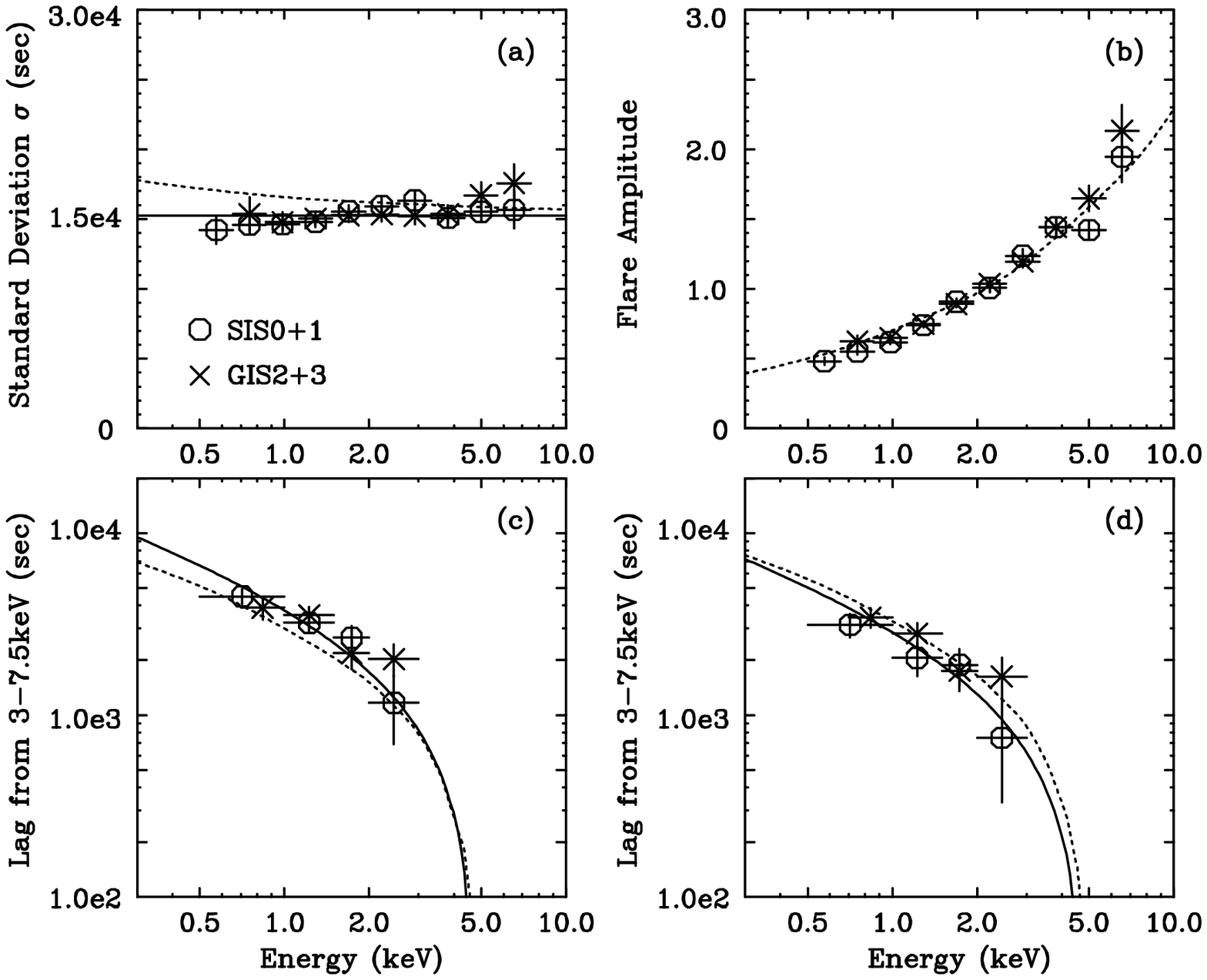]{Measurement of the source parameters for a homogeneous
synchrotron model describing the 1994 May flare 
of PKS 2155$-$304. The circles are combined SIS data, while 
the crosses are combined GIS data.
(a): The flare duration, estimated by the standard deviation, $\sigma$, 
of the Gaussian fits at various X--ray energies 
(see $\S$ 2.2.1). 
The solid line is a constant fit ($\sigma$ = 1.5 $\times$ 10$^{4}$ sec), 
while the dashed line is calculated from the Gaussian fit of the model 
light curves in $\S$ 3.
(b): The flare amplitude ($Ap$ $\equiv$ $C_{1}$/$C_{0}$) at various 
X--ray energies, determined by a Gaussian fit in $\S$ 2.2.1. 
The dashed line is calculated from the Gaussian fit of the modeled 
light curves in $\S$ 3.
(c): Time lag of photons of various X--ray energies versus the 3.0 $-$ 
7.5 keV band photons, calculated from the peaking time determined by a 
Gaussian fit. The solid line is the best fit with a function given in 
$\S$ 2.2.1 ($B$ = 0.11 G, $\delta$ = 25). The dashed 
line is calculated from the Gaussian fit of the model light curves in $\S$ 3.
(d): Time lag of photons of various X--ray energies versus the 3.0 $-$ 
7.5 keV band photons, calculated using the discrete correlation function 
(DCF). 
The solid line is the best fit with a function given in $\S$ 2.2.1 
($B$ = 0.13 G, $\delta$ = 25). The dashed line is 
as in (c).
\label{fig.5}}

\figcaption[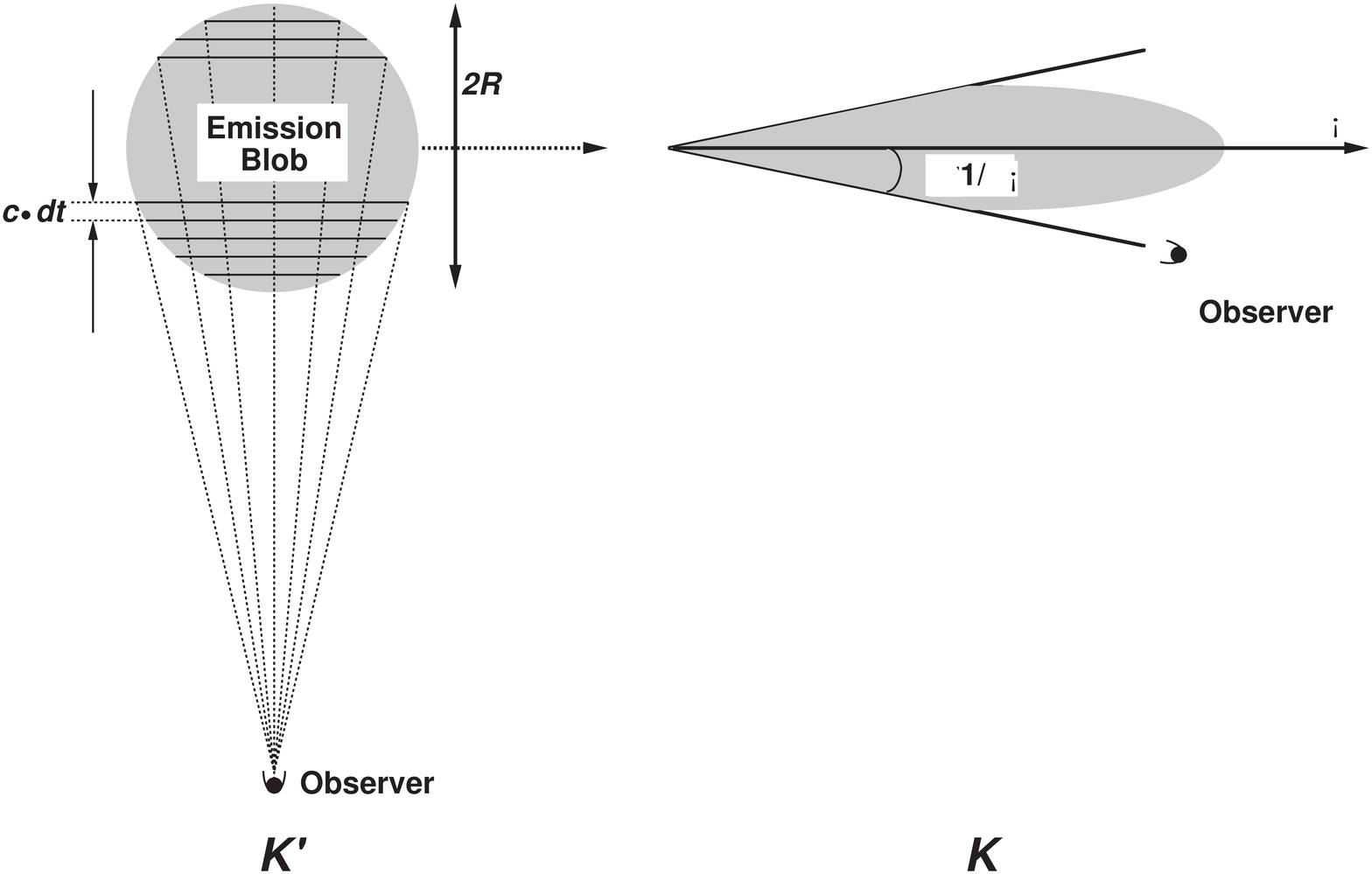]{The schematic view of the division of the emission 
blob into 'slices'. Left Panel: Source frame ($K'$). Right Panel: Observer's 
frame ($K$). The emission blob is first cut into slices in the source 
frame and contributions from each shell are summed. In the observer's 
frame, the radiation is concentrated in a narrow cone with a half angle 
$\theta$ $\simeq$ 1/$\Gamma$. \label{fig.6}}

\figcaption[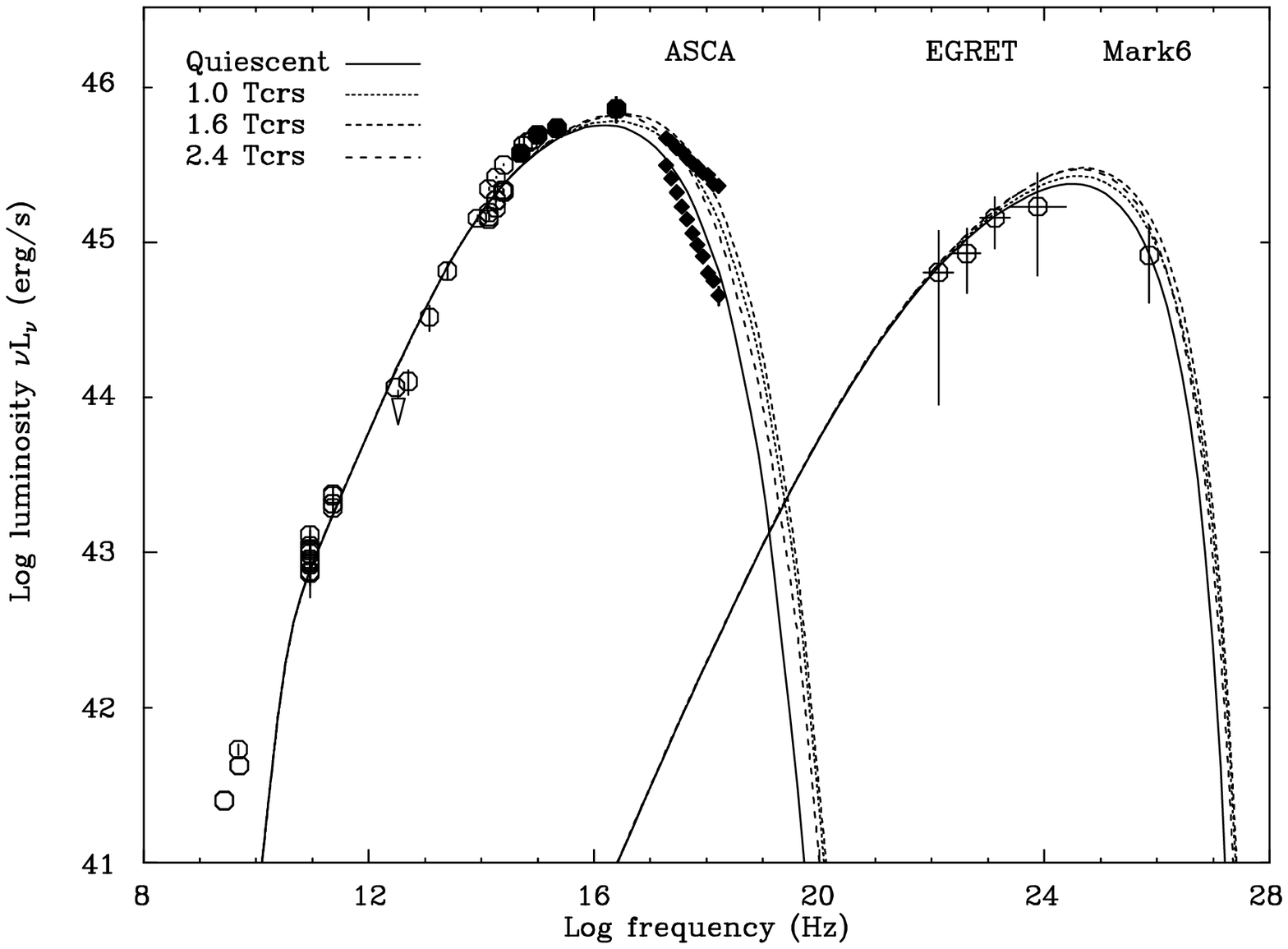]{Multi-band spectrum of PKS 2155$-$304. The filled circles
are the nearly simultaneous \iue and \euve data reported in Urry et al. 
(1997). The filled squares are the X--ray data obtained with \asca 
(this work). The data from both GISs are combined.
The high state \asca data are from 
the 15 ksec of the peak flux of the flare, while the low state data are
from the 15 ksec near the end of the observation. \egret data and TeV data 
are from Vestrand, Stacy, \& Sreekumar (1995) and Chadwick et al. (1999), 
respectively. 
The other non-simultaneous data are from the NED data base. The solid line 
and dashed lines represent the time evolution of the photon spectra 
calculated from the time-dependent SSC model described in $\S$ 3. The 
physical (model) parameters are given in Table 2. 
The multi-frequency spectra at $t$/$t_{\rm crs}$ = 
0, 1.0, 1.6 and 2.4 are shown respectively.\label{fig.7}}

\figcaption[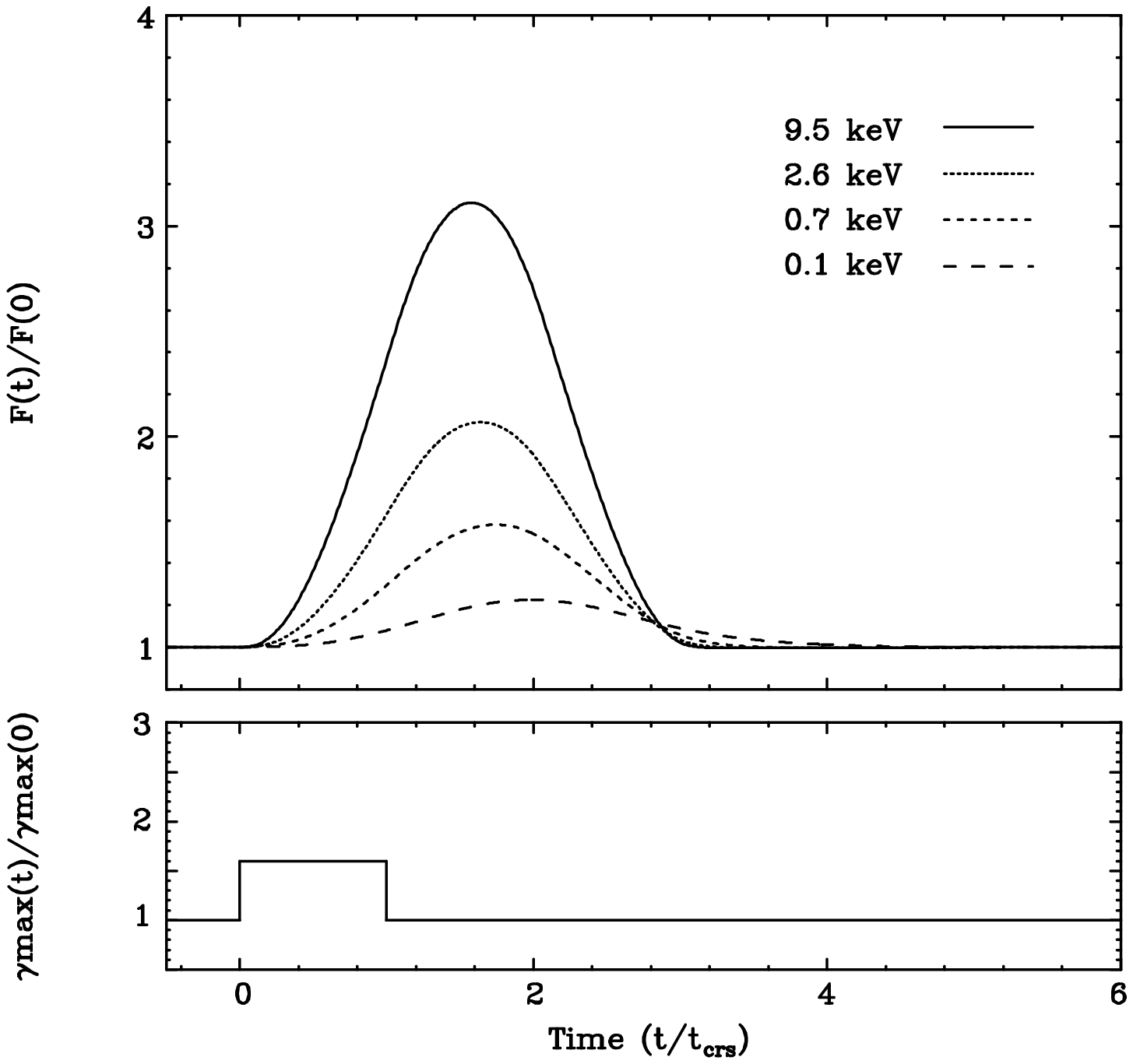]{Upper Panel: Simulated light curves at different 
UV/X--ray energies, reproducing the rapid flare of PKS 2155$-$304 in 1994 
May. We plotted the time evolution of the fluxes at different frequencies, 
normalized to the initial ($t$ = 0) value. 
The quasi-symmetric shape of the high energy light curves and the increasing
time lag of the peak with decreasing energy is clearly seen. 
Lower Panel: Assumed change of $\gamma_{\rm max}$ during the flare. The 
$\gamma_{\rm max}$ is assumed to increase by a factor of 1.6 from the 
initial value for 0 $\le$ $t$ $\le$ $t_{\rm crs}$. 
Time-axes of both panels are in units of the source light-crossing time 
($t_{\rm crs}$). 
In the observer's frame, $t_{\rm crs}$ corresponds to 0.3 day. 
\label{fig.8}}
\clearpage

\begin{deluxetable}{crrr}
\footnotesize
\tablecaption{Input Observables for PKS 2155$-$304}
\tablewidth{0pt}
\tablehead{
\colhead{Parameters} & \colhead{Input Values}}
\startdata
$\nu_{\rm s,17}$ (in 10$^{17}$ Hz) & 0.3 
\nl
$\nu_{\rm c,27}$ (in 10$^{27}$ Hz) & 0.3
\nl
$\nu_{\rm br,14}$ (in 10$^{14}$ Hz) & 0.5 
\nl
$t_{5}$ (in 10$^{5}$ sec) & 0.3 
\nl
$f_{\rm sync}$ (in 10$^{-10}$ erg cm$^{-2}$ s$^{-1}$) & 13
\nl
$f_{\rm SSC}$ (in 10$^{-10}$ erg cm$^{-2}$ s$^{-1}$) & 4.1
\nl
$\alpha$ & \hspace{10mm}0.18 
\nl
\enddata
\tablenotetext{}{Input parameters (observational quantities) determined from 
the multi-frequency spectrum of PKS 2155$-$304 (Figure 7), as defined
in \S 3.2.\\
}\end{deluxetable}

\begin{deluxetable}{crrrrrrrrrr}
\footnotesize
\tablecaption{Output Physical Quantities for PKS 2155$-$304}
\tablewidth{0pt}
\tablehead{
\colhead{Parameters}& \colhead{Relation to Observables} & 
\colhead{Output Values}}
\startdata
$\delta$ & 3.0$\times$10$^{-6}$$d_{L}^{1/4}$$f_{\rm sync}^{1/4}$$f_{\rm SSC}^{-1/
8}$$\nu_{\rm s,17}^{-1/4}$$\nu_{\rm c,27}^{1/2}$$t_{5}^{-1/4}$& 28 
\nl
$B$ (G) & 3.8$\times$10$^{-9}$$d_{L}^{1/4}$$f_{\rm sync}^{1/4}$$f_{\rm SSC}^{-1/8
}$$\nu_{\rm s,17}^{3/4}$$\nu_{\rm c,27}^{-3/2}$$t_{5}^{-1/4}$ & 0.14 
\nl
$R$ (cm)& 8.8$\times$10$^{9}$$d_{L}^{1/4}$$f_{\rm sync}^{1/4}$$f_{\rm SSC}^{-1/8}
$$\nu_{\rm s,17}^{-1/4}$$\nu_{\rm c,27}^{1/2}$$t_{5}^{3/4}$ & 2.4$\times$10$^{16}
$ 
\nl
$t_{\rm esc}$ (sec) & 6.3$\times$10$^{14}$$d_{L}^{-1/4}$$f_{\rm sync}^{-1/4}$$f_{
\rm SSC}^{1/8}$(1+$f_{\rm SSC}$/$f_{\rm sync}$)$^{-1}$$\nu_{\rm s,17}^{-5/4}$$\nu
_{\rm c,27}^{5/2}$$\nu_{\rm br,14}^{-1/2}$$t_{5}^{1/4}$ & 8.6$\times$10$^{6}$
\nl
$\gamma_{\rm max}$ & 2.7$\times$10$^{12}$$d_{L}^{-1/4}$$f_{\rm sync}^{-1/4}$
$f_{\rm SSC}^{1/8}$$\nu_{\rm s,17}^{1/4}$$\nu_{\rm c,27}^{1/2}$$t_{5}^{1/4}$ & 8.3$\times$10$^{4}$ 
\nl 
$q_e$ (cm$^{-3}$ s$^{-1}$) & --- (normalized to agree with $f_{\rm sync}$)&6.1$\times$10$^{-7}$
\nl 
$s$ & 2$\alpha$+1 & 1.35 
\nl 
\enddata
\tablenotetext{}{Output model parameters for the multi-frequency 
spectrum of PKS 2155$-$304 (Figure 7). $d_L$ = 1.47$\times$10$^{27}$ cm
is the luminosity distance of PKS 2155$-$304. The emission 
region size $R$ is approximated as $R$ $\sim$ $c$$t_{\rm var}$$\delta$. 
We adopted a cutoff power law for the injected electron 
population 
: $Q(\gamma)$ = $q_e$ $\gamma^{-s}$ exp($-$$\gamma$/$\gamma_{\rm max}$). 
We set the minimum Lorentz factor of the electrons, $\gamma_{\rm min}$ = 1,  
for simplicity.
}\end{deluxetable}

\clearpage

\begin{figure}
\epsscale{1.0}
\plotone{f1.eps}
\end{figure}
\clearpage

\begin{figure}
\epsscale{1.0}
\plotone{f2.eps}
\end{figure}
\clearpage

\begin{figure}
\epsscale{1.0}
\plotone{f3.eps}
\end{figure}
\clearpage

\begin{figure}
\epsscale{1.0}
\plotone{f4.eps}
\end{figure}
\clearpage

\begin{figure}
\epsscale{1.0}
\plotone{f5.eps}
\end{figure}
\clearpage

\begin{figure}
\epsscale{1.0}
\plotone{f6.eps}
\end{figure}
\clearpage

\begin{figure}
\epsscale{1.0}
\plotone{f7.eps}
\end{figure}
\clearpage

\begin{figure}
\epsscale{1.0}
\plotone{f8.eps}
\end{figure}
\clearpage

\end{document}